\pdfoutput=1

\documentclass{article}
\usepackage{icmc2022template}
\usepackage{times}
\usepackage{ifpdf}
\usepackage{soul}
\usepackage[english]{babel}

\def\papertitle{Seeing Sounds, Hearing Shapes: a gamified study to evaluate sound-sketches}
\def\firstauthor{Sebastian Löbbers}
\def\secondauthor{György Fazekas}

\newif\ifpdf
\ifx\pdfoutput\relax
\else
   \ifcase\pdfoutput
      \pdffalse
   \else
      \pdftrue
  \fi
\fi

\ifpdf 
  \usepackage[pdftex,
    pdftitle={\papertitle},
    pdfauthor={\firstauthor, \secondauthor},
    bookmarksnumbered, 
    pdfstartview=XYZ 
   ]{hyperref}

  \usepackage[pdftex]{graphicx}
  \graphicspath{{./figures/}}
  \DeclareGraphicsExtensions{.pdf,.jpeg,.png}

  \usepackage[figure,table]{hypcap}

\else 
  \usepackage[dvips,
    bookmarksnumbered, 
    pdfstartview=XYZ 
  ]{hyperref}  

  \usepackage[dvips]{epsfig,graphicx}
  \graphicspath{{./figures/}}
  \DeclareGraphicsExtensions{.eps}

  \usepackage[figure,table]{hypcap}
\fi

\hypersetup{
    colorlinks,%
    citecolor=black,%
    filecolor=black,%
    linkcolor=black,%
    urlcolor=black
}

\title{\papertitle}

 \twoauthors
   {0.5in}
   {\firstauthor} {Centre for Digital Music \\ Queen Mary University of London \\ %
   {\tt \href{mailto:s.lobbers@qmul.ac.uk}{s.lobbers@qmul.ac.uk}}}
   {\secondauthor} {Centre for Digital Music \\ Queen Mary University of London \\ %
     {\tt \href{mailto:m.barthet@qmul.ac.uk}{g.fazekas@qmul.ac.uk}}}

\begin{document}
\sloppy
\capstartfalse
\maketitle
\capstarttrue
\begin{abstract}
Sound-shape associations, a subset of cross-modal associations between the auditory and visual domain, have been studied mainly in the context of matching a set of purposefully crafted shapes to sounds. Recent studies have explored how humans represent sound through free-form sketching and how a graphical sketch input could be used for sound production. In this paper, the potential of communicating sound characteristics through these free-form sketches is investigated in a gamified study that was conducted with eighty-two participants at two online exhibition events. The results show that participants managed to recognise sounds at a higher rate than the random baseline would suggest, however it appeared difficult to visually encode nuanced timbral differences.    
\end{abstract}

\section{Introduction}\label{sec:introduction}

Sample libraries and sound synthesisers are ubiquitous in modern digital music production. However, with samples often organised in lists, finding the ``right'' sound can be a tedious, frustrating task and, similarly, complex synthesiser controls make it difficult to realise sound ideas in a straightforward way. Tagging of sound material and synthesiser pre-sets improves the search process, but can be imprecise because of the ambiguity of language used to describe sound, or more specifically musical timbre \cite{saitisTimbreSemanticsLens2018,siedenburgComparisonApproachesTimbre2016,mcadamsPerceptionMusicalTimbre2016}. Sound visualisations can be an additional aid for communicating a sample's characteristics to a user as seen with waveform representations in digital audio workstations (DAWs) or spectrograms in specialised audio software like Izotope’s \textit{RX} and \textit{Iris}. Further, cross-modal associations between the visual and auditory domain, for example between colours or shapes and sound, can be harnessed for query or perceptually informed visualisations. This paper presents a study that investigates to what extent sound characteristics can be communicated through graphical sketches of personal cross-modal associations. Participants were presented with a gamified survey in which they had to match sounds with sketches that were created in a different study by participants who will be referred to as artists in this paper. The work was conducted in the context of a larger research programme that investigates how sketches can be used to drive sound synthesis~\cite{lobbers2021buildingsynth}. This section covers relevant research into sound-shape associations and sound visualisations and briefly introduces gamified study design principles. The rest of the paper is organised as follows: Sections \ref{sec:methods}, \ref{sec:analysis} and \ref{sec:results} describe the design, analysis and results of the study, a discussion and conclusion are provided in Sections \ref{sec:discussion} and \ref{sec:conclusion}.

\subsection{Background on sound-shape associations and sound visualisation}\label{subsec:related_work} 
Research suggests that most humans associate sound with elements of the visual domain to some extent~\cite{martinoSynesthesiaStrongWeak2001}. This paper focuses on cross-modal associations between sounds and shapes that were first researched by Wolfgang K{\"o}hler who found that the made-up words \textit{takete} and \textit{maluma}\footnote{Other studies use \textit{bouba} and \textit{kiki}.} are associated with jagged and round shapes. The effect was observed across cultures~\cite{davisFitnessNamesDrawings1961,taylorPhoneticSymbolismFour1962,bremnerBoubaKikiNamibia2013}, age groups including toddlers~\cite{maurerShapeBoubasSound2006}, for musical sounds~\cite{adeliAudiovisualCorrespondenceMusical2014} and in a recent study where participants were asked to represent musical sounds through sketching~\cite{lobbers2021sketching}. Sound-shape associations can inform visualisations that communicate sound characteristics without audio playback for example to improve the organisation of one's personal music library~\cite{kolhoffMusicIconsProcedural2006}, sample selection for live DJ performances~\cite{chen2010thumbnaildj}, exploration of different natural sounds~\cite{wan2019towards,ishibashi2020investigating} and retrieval of abstract sounds ~\cite{grillVisualizationPerceptualQualities2012}. The idea to use sketches as an input for sound production was first explored with a non-functioning prototype~\cite{kneesSearchingAudioSketching2016}. Further development~\cite{engeln2021similarity} was made possible through current advancements in deep learning for sketch recognition that is informed by the release of the large-scale sketch dataset \textit{Quick, Draw!}~\cite{haNeuralRepresentationSketch2017}.\footnote{\href{https://quickdraw.withgoogle.com/data}{quickdraw.withgoogle.com/data}} This study investigates to what extent human participants can recognise sounds from simple sketches to identify perceptually informed features that can help build a deep learning pipeline for sketch-based sound production tasks. 

\subsection{Study gamification}\label{subsec:gamification}
This study was designed with gamification principles in mind. Research suggests that gamification can have a positive effect on quality and quantity of responses~\cite{hamariDoesGamificationWork2014} by incorporating the motivational elements of games~\cite{rappStrengtheningGamificationStudies2019}. For example, participation can be incentivised by a point system~\cite{harmsGamificationOnlineSurveys2015} with the option to compare results with other ``players''\footnote{See \href{https://www.themusiclab.org/}{Harvard's \textit{Musical IQ} survey} for an example.} or by an interface design that adopts a game-like aesthetic as it can be seen in \textit{The Clapping Game}~\cite{duffy2018makes} and \textit{Microjam}~\cite{martin2020data}. The effectiveness of the design can be evaluated through user experience survey questions as proposed by Morschheuser et al.~\cite{morschheuserHowGamifyMethod2017}.  

\section{Methods and Material}\label{sec:methods}
This section describes the design of the study. In ten rounds, one sound and four different sketches were presented of which only one sketch was created with that sound in mind. Participants were tasked to find that match. Each round featured a different sound and sketches from a different artist. The following hypotheses are investigated:
\begin{itemize}
    \item On average, participants will find matches at a higher rate than the random baseline.
    \item Participants are most likely to select the sketch corresponding to the next most similar sound when not selecting the correct match.
    \item The type of sound and an artist's representational sketch style will have an influence on the match rate. The match rate is defined as the frequency at which the correct sound-sketch match is selected.
\end{itemize}

\subsection{Participants}\label{subsec:participants}
Eighty-two participants took part in the study of which thirty completed an optional post-study survey. The study was part of two online exhibitions that were presented at the Ars Electronica Festival 2020\footnote{\href{https://forkingpaths.intersections.io/}{\textit{The Garden of Forking Paths}} at the Ars Electronica Festival 2020} and the Edinburgh Science Festival 2021.\footnote{\href{https://www.seeingmusic.app/}{\textit{Seeing Music}} at the Edinburgh Science Festival 2021} Sixteen participants completed the study on a mobile device (9 iPhone, 6 Android, 1 iPad) and sixty-six used a desktop or laptop computer (32 Mac, 25 Windows, 5 Linux, 4 unknown). Of the participants who completed the survey 13 were female, 16 male and 1 did not disclose this information. The mean age was 33.6 years ($\sigma$ = 11.3). Fourteen stated to actively engage in musical activity at least multiple times a month and two reported a visual impairment (1 short-sighted, 1 near-sighted) and one an auditory impairment (tinnitus).

\subsection{Stimuli}\label{subsec:stimuli}
All ten sound stimuli and sketches from eleven different artists were selected from a previous study \cite{lobbers2021sketching} where participants were asked to sketch their personal associations with the sound stimuli. An artist was selected if none of their sketches contained any direct reference to the sound such as depictions of instruments or other recognisable sound sources or symbolic representations like letters or icons. All sketches were created with a MacBook touchpad on a monochromatic digital sketch interface with fixed stroke width. Sounds are equally pitched, loudness normalised and range from musical instruments (\textit{Piano}, \textit{Strings}, \textit{Electric Guitar}) and environmental sounds (\textit{Impact}) to synthesised pads (\textit{Telephonic}, \textit{Subbass}) and abstract textures (\textit{Noise}, \textit{String Grains}, \textit{Crackles}, \textit{Processed Guitar}).\footnote{All sounds can be accessed online together with the sketches drawn by participants during the experiment \url{https://bit.ly/3ta6crU}.}

\subsection{Apparatus}\label{subsec:apparatus}
The study was implemented as a web app in \textit{p5.js} with mobile and desktop capability. The interface was designed with a gamified aesthetic in mind following examples introduced in Section \ref{subsec:gamification}. The study consisted of ten rounds. In each round, participants were presented with one sound and four sketches of which one was created with that sound in mind and the remaining three in reference to the 3\textsuperscript{rd}, 6\textsuperscript{th} and 9\textsuperscript{th} most dissimilar sound as described in Section \ref{subsec:dissimilarity}. Sketches were re-drawn in real-time during audio playback following their normalised timestamp data. Participants were asked to find the correct match and after selection, the interface revealed whether their choice was correct. The order of the sounds and artists were randomised on each run. Cookies were used to track returning participants and prevent multiple submissions from the same device. 

\begin{figure}[h]
\centering
\vspace*{-0.1in}
\includegraphics[width=0.9\columnwidth]{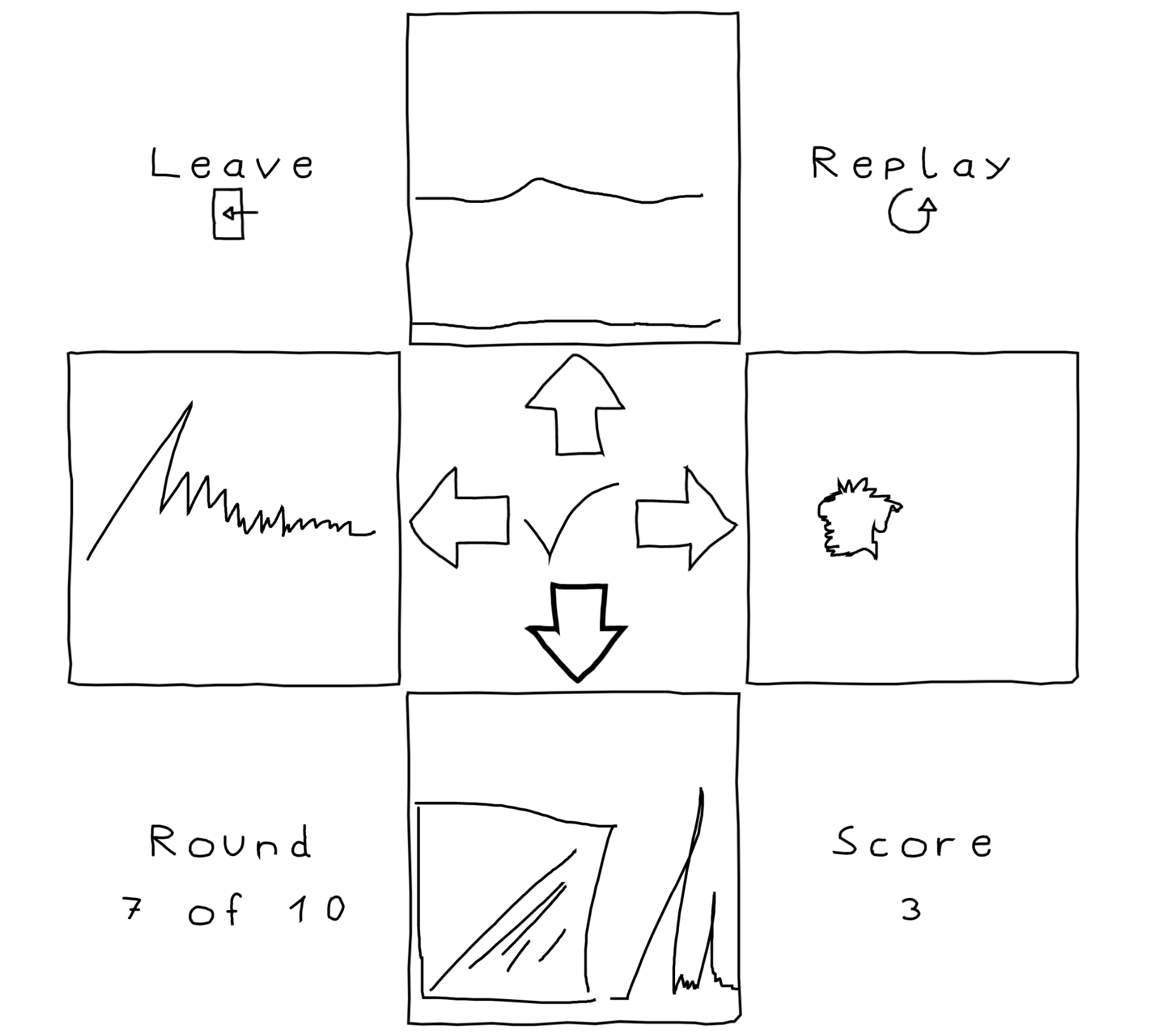}
\vspace*{-0.1in}
\caption{Screenshot of the study interface. The study can be accessed online at \url{https://phd-studies-eddd5.web.app/}.}
\label{fig:Interface}
\end{figure}

\subsection{Procedure}\label{subsec:procedure}
In order to make the study more accessible, participants were presented with only one introduction page that included a short description and an animated image to explain how the game works before starting the matching task. Only after completing all rounds, participants were asked for consent to save their data. To incentivise submission, participants could compare their score to other participants after their data had been saved using a mechanism similar to the \textit{Musical IQ} survey described in Section \ref{subsec:gamification}. Additionally, participants were invited to take part in an optional survey that collected basic personal information (age, gender, visual or auditory impairment), music proficiency using a subset of the GoldMSI survey framework \cite{mullensiefenGoldsmithsMusicalSophistication} and their experience with the study interface. Information about the device used to complete the study were collected automatically. 

\section{Analysis}\label{sec:analysis}
This section provides an overview of the quantitative and qualitative methods used to analyse the study data.

\subsection{Sound stimuli dissimilarity}\label{subsec:dissimilarity}
One hypothesis introduced in Section \ref{sec:methods} states that participants are most likely to select the sketch corresponding to the next most similar sound when not selecting the correct match. To investigate this, the four sketches in each round relate to the played sound and the 3\textsuperscript{rd}, 6\textsuperscript{th} and 9\textsuperscript{th} most similar sound. To quantify dissimilarity, the euclidean distance between standardised audio feature vectors was calculated similar to \cite{grillVisualizationPerceptualQualities2012}. Audio features were extracted with the Librosa library and timbre models \cite{pearceModellingTimbralHardness2019} as described in \cite{lobbers2021sketching}.\footnote{A detailed summary of audio feature extraction can be found at \url{http://doi.org/10.5281/zenodo.4764351}}

\begin{figure}[h]
\centering
\vspace*{-0.15in}
\includegraphics[width=0.9\columnwidth]{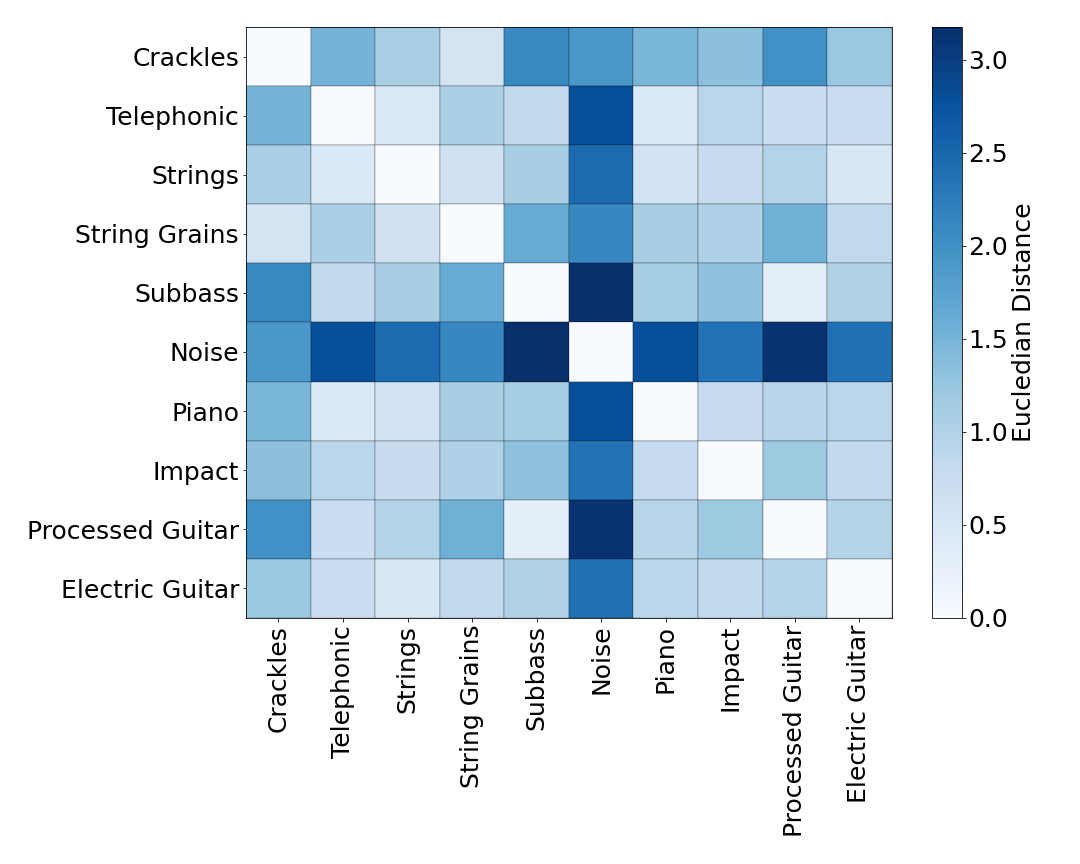}
\vspace*{-0.1in}
\caption{Dissimilarity matrix between sound stimuli. A darker shade indicates a larger euclidean distance between standardised audio features. \label{fig:dissimilarity}}
\end{figure}

\subsection{Survey response analysis}\label{subsec:survey_analysis}
In addition to general personal information, the survey contains questions asking whether participants enjoyed the study design. Taking an excerpt from Harms et al.'s survey questions \cite{harmsGamificationOnlineSurveys2015}, responses were collected using a five point Likert scale on whether they enjoyed playing the game and whether they found the task difficult. In addition, participants were given the opportunity to report any issues that they might have encountered and add any general comments. Likert scale responses were summarised and free-form responses were analysed thematically \cite{braunUsingThematicAnalysis2006}. 

\subsection{Statistical analyses}\label{subsec:statistical}
The first hypothesis introduced in Section \ref{sec:methods} states that on average participants will select correct matches at a higher rate than the random baseline of 2.5 out of 10. Pearson's Chi-squared test was used to test for significant differences between the collected scores and the random distribution. 

Another hypothesis postulates that participants will be more likely to select the sketch of the next most similar sound than the one of the most dissimilar. For this, the match rate for each sound was extracted from the data. However, since the artist and sound order was randomised independently of each other, artists are not evenly distributed among all sounds. Calculating a sound's match rate by dividing the total number of matches by the total number of occurrences might introduce biases because of an artist's over-representation. This was mitigated by calculating the match rate for each individual sound-sketch and ranking them from 10 (highest rate) to 1 (lowest rate) within each artist to obtain the mean rank for each sound. The same procedure was used to compare match rates between artists.

Spearman's rank correlation was used to investigate whether the mean ranks for sounds correlate with their mean dissimilarity quantified by the mean euclidean distance described in Section \ref{subsec:dissimilarity} and whether the mean rank for artists correlates with their sketching style. The sketching style of an artist was quantified by counting the frequency of each of the five categories \textit{Grains}, \textit{Lines}, \textit{Object/Scenes}, \textit{Chaotic/Jagged}, \textit{Radiating} described in \cite{lobbers2021sketching}, however because of the artist filtering described in Section \ref{subsec:stimuli} the \textit{Object/Scenes} category was removed almost entirely from the subset.

\section{Results}\label{sec:results}
The outcome of statistical analysis of the matching task and results of the post-study survey are reported in this section. 

\subsection{Overall match rate}\label{subsec:hitrate}
\begin{figure}[h]
\centering
\vspace*{-0.15in}
\includegraphics[width=0.9\columnwidth]{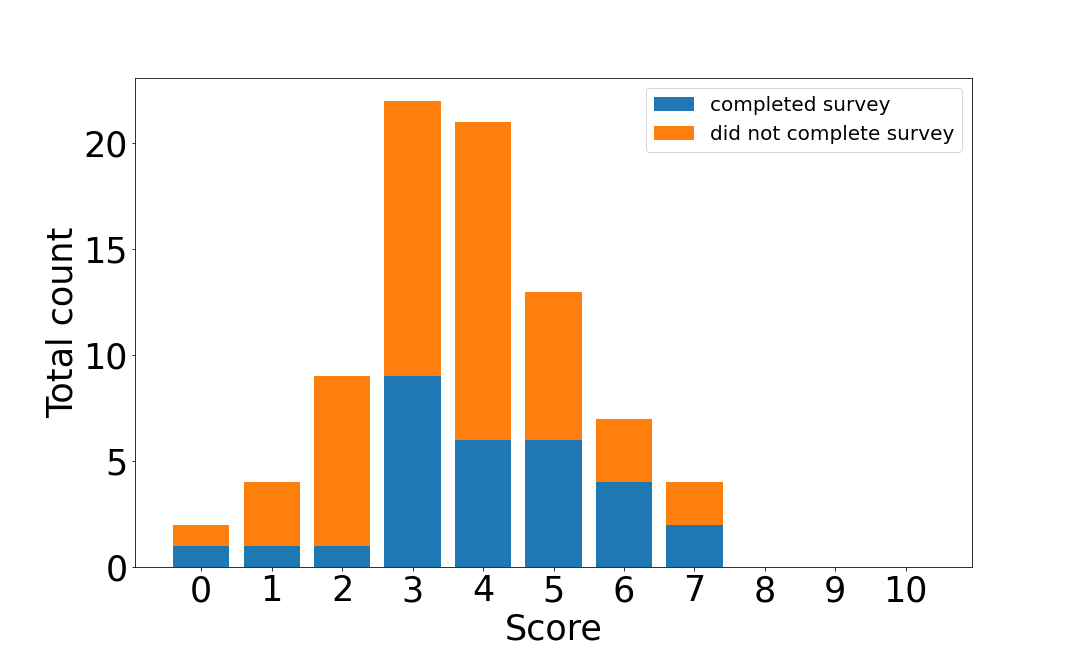}
\vspace*{-0.1in}
\caption{Match rate of all participants. The mean match rate is 3.74 for all participants, 4.07 for participants who completed the survey and 3.56 for participants who did not complete the survey. 
}
\label{fig:scores}
\end{figure}

Figure \ref{fig:scores} shows that participants mostly found three (22 participants), four (21 participants) or five (13 participants) correct matches. Chi-square test returned a significant difference between this distribution and the random baseline described in Section \ref{subsec:statistical} (${\chi}^2$(1,~N~$=$~82)~$=$~99509.11 p~$<$~.0001). In total, the correct option was selected most frequently (307/820 selections), however the second most frequent selection was the option that refers to the most dissimilar sound (212/820). These results suggest that participants were able to extract information from the sketches that helped them to find correct matches, however, contrary to expectation, participants did not predominately choose the sketch referring to the next most similar sound when not finding the correct match.

\subsection{Match rate by sound}\label{subsec:hitrate_sound}

\begin{figure}[h]
\centering
\vspace*{-0.15in}
\includegraphics[width=0.9\columnwidth]{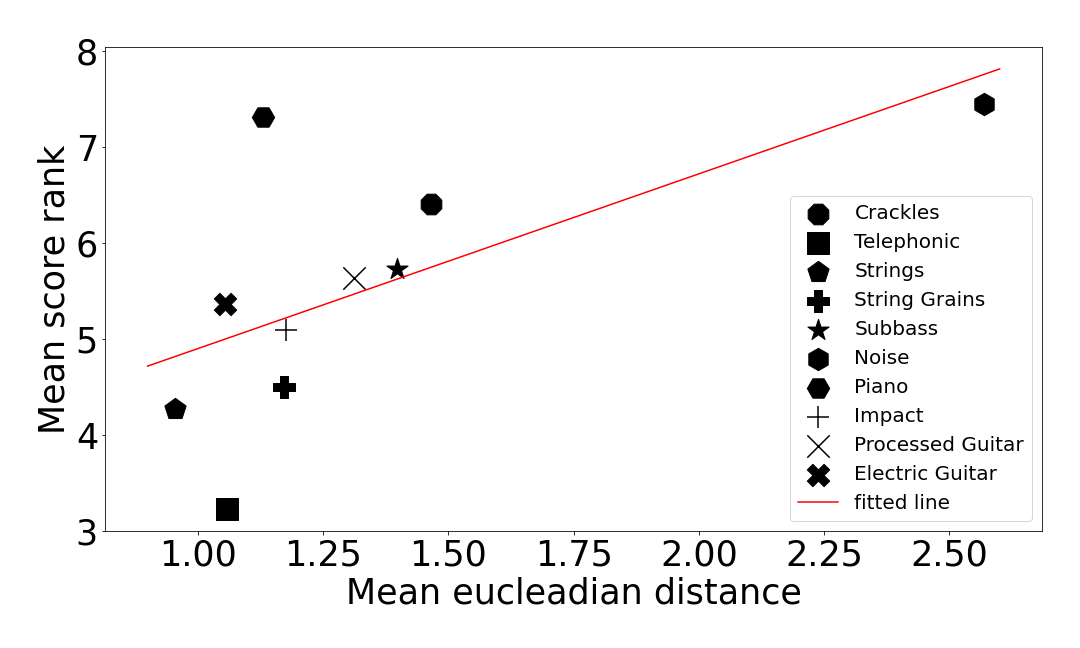}
\vspace*{-0.1in}
\caption{Correlation between match rate rank by sound and sound dissimilarity quantised by the mean euclidean distance between audio features.}
\label{fig:correlation_sound}
\end{figure}

Table \ref{tab:selection_table} shows how often each of the four sketch options were selected for each sound. Except for \textit{Telephonic} (18/82 selections) and \textit{String Grains} (22/82) the correct option was selected most frequently with \textit{Noise} (49/82) and \textit{Piano} (47/82) showing the highest match rate. Figure \ref{fig:dissimilarity} shows that with an average euclidean distance of 2.57 \textit{Noise} is the most dissimilar sound to all other sounds except for \textit{Crackles} which is the second most dissimilar sound with a euclidean distance of 1.46. These sounds also show a high mean match rank  7.45 for \textit{Noise} ($\sigma$ = 2.96) and 6.4 for \textit{Crackles} ($\sigma$ = 2.27) compared to sounds that are less dissimilar like \textit{Telephonic} with a euclidean distance of 1.06 and a mean rank of 3.23 ($\sigma$ = 1.95) or \textit{Strings} with a euclidean distance of 0.96 and a mean rank of 4.27 ($\sigma$ = 2.89). Spearman rank correlation suggests a strong positive correlation between the mean dissimilarity of a sound and its mean score rank ($\rho$ $=$ .7, p $<$ .05) as visualised with a trend line in Figure \ref{fig:correlation_sound}. However, the figure also shows that \textit{Piano} has the second highest mean rank ($\mu$ = 7.32, $\sigma$ = 2.25) despite a mid-range dissimilarity (euclidean distance of 1.13). These results suggest that some sketches are more likely to be correctly matched to a sound which is at least partially influenced by the sound's mean dissimilarity to the remaining stimuli.

\begin{table*}[t]
\begin{center}
    \begin{tabular}{|llll|}
\hline
Correct Option  & 2. Option (3\textsuperscript{rd} distance) & 3. Option (6\textsuperscript{th} distance) & 4. Option (9\textsuperscript{th} distance) \\
               
\hline
Crackles: 30         &      Electric Guitar: 10 &           Telephonic: 24 &              Subbass: 18 \\
Telephonic: 18       &     Processed Guitar: 22 &               Impact: 18 &                Noise: 24 \\
Strings: 28          &                Piano: 14 &     Processed Guitar: 24 &                Noise: 16 \\
String Grains: 22    &      Electric Guitar: 13 &                 Piano: 4 &                Noise: 43 \\
Subbass: 27          &      Electric Guitar: 15 &               Impact: 21 &                Noise: 19 \\
Noise: 49            &               Impact: 13 &            Telephonic: 4 &              Subbass: 16 \\
Piano: 47            &               Impact: 17 &         String Grains: 7 &                Noise: 11 \\
Impact: 32           &      Electric Guitar: 19 &     Processed Guitar: 11 &                Noise: 20 \\
Processed Guitar: 29 &                 Piano: 7 &               Impact: 24 &                Noise: 22 \\
Electric Guitar: 25  &               Impact: 23 &     Processed Guitar: 11 &                Noise: 23 \\
\hline
\textbf{Total: 307} & \textbf{Total: 153} & \textbf{Total: 148} & \textbf{Total: 212}\\
\hline
\end{tabular}
\end{center}
\vspace*{-0.05in}
 \caption{Match rates for each sound. For most sounds the correct option was selected most of the times. However, some sounds were often mismatched (e.g. String Grains and Noise).\label{tab:selection_table}}
 
\end{table*}

\begin{table*}[t]
 \begin{center}
 \begin{tabular}{|p{0.06\linewidth}||p{0.06\linewidth}|p{0.06\linewidth}|p{0.06\linewidth}|p{0.06\linewidth}|p{0.06\linewidth}|p{0.06\linewidth}|p{0.06\linewidth}|p{0.06\linewidth}|p{0.06\linewidth}|p{0.06\linewidth}|}

    \hline
    
    \textbf{Sound Name}
    &
    Crackles
    &
    Tele-phonic
    &
    Strings
    &
    String Grains
    &
    Subbass
    &
    Noise
    &
    Piano
    &
    Impact
    &
    Proc. Guitar
    &
    Electric Guitar
    \\
  \hline
  \hline
    Best
    &
    \includegraphics[width=1.05\linewidth]{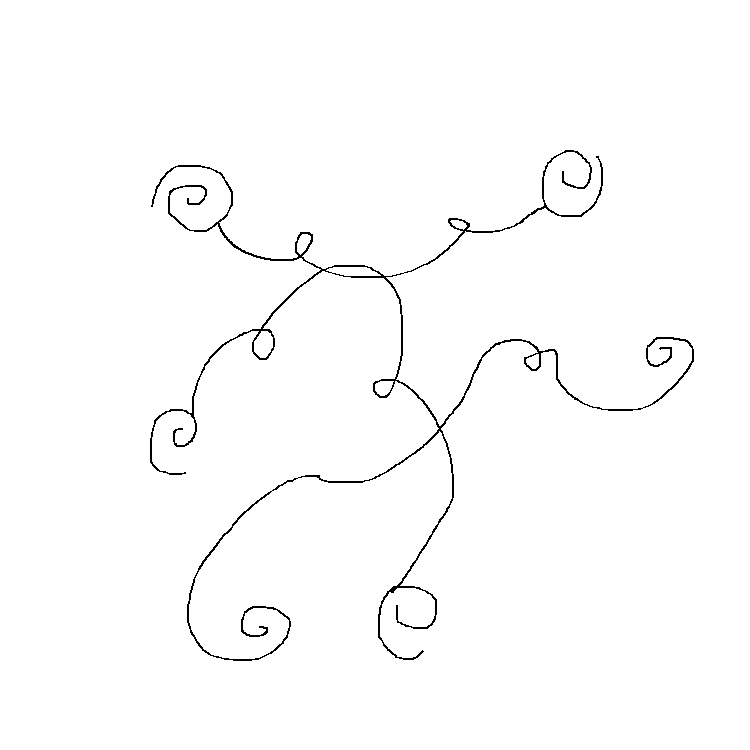} 
    6/11
    &
    \includegraphics[width=1.05\linewidth]{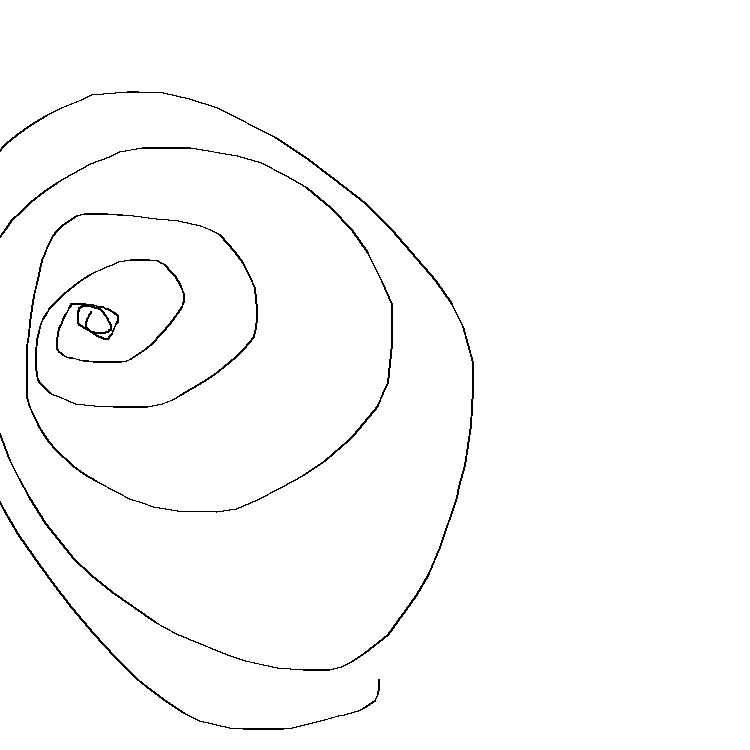}
    \underline{7/12}
    &
    \includegraphics[width=1.05\linewidth]{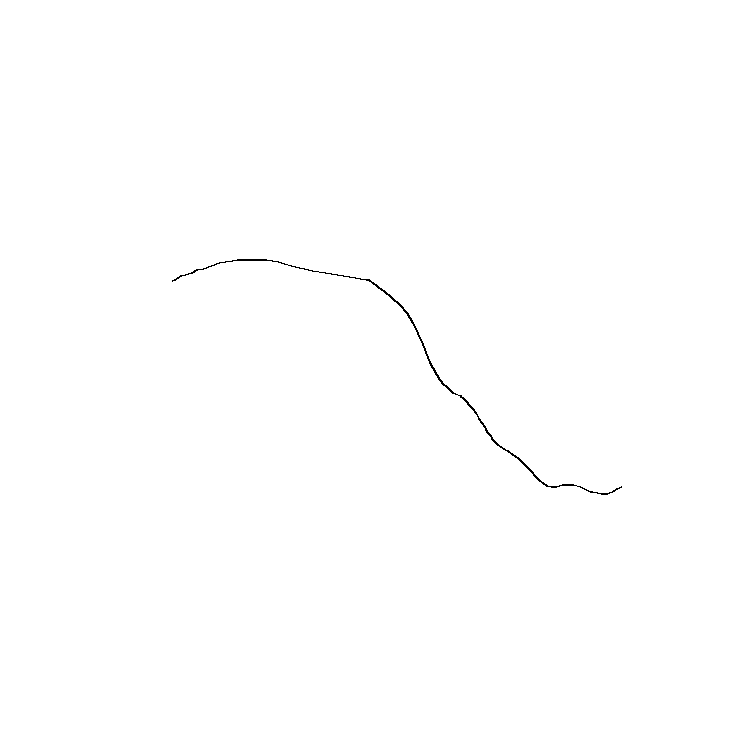}
    9/15
    &
    \includegraphics[width=1.05\linewidth]{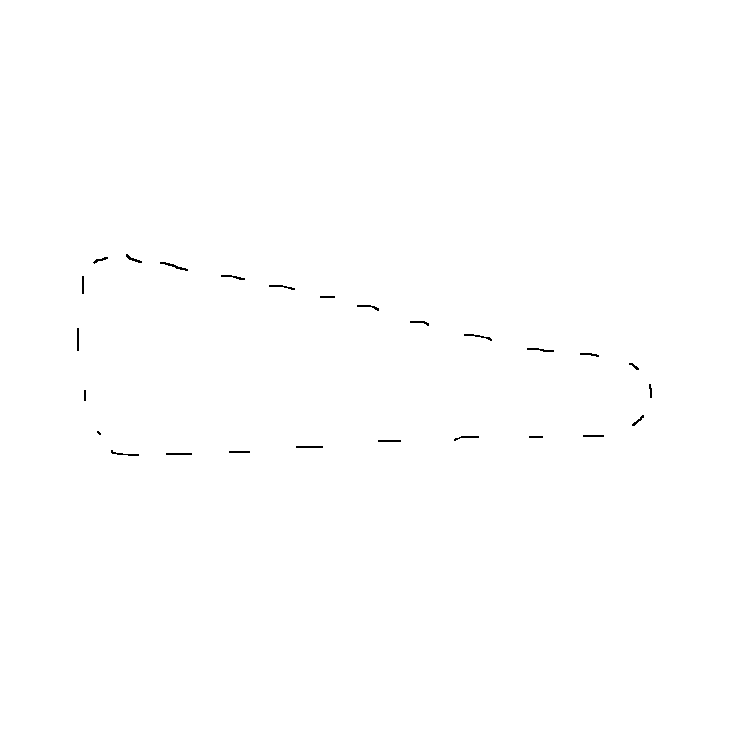}
    \underline{4/9}
    &
    \includegraphics[width=1.05\linewidth]{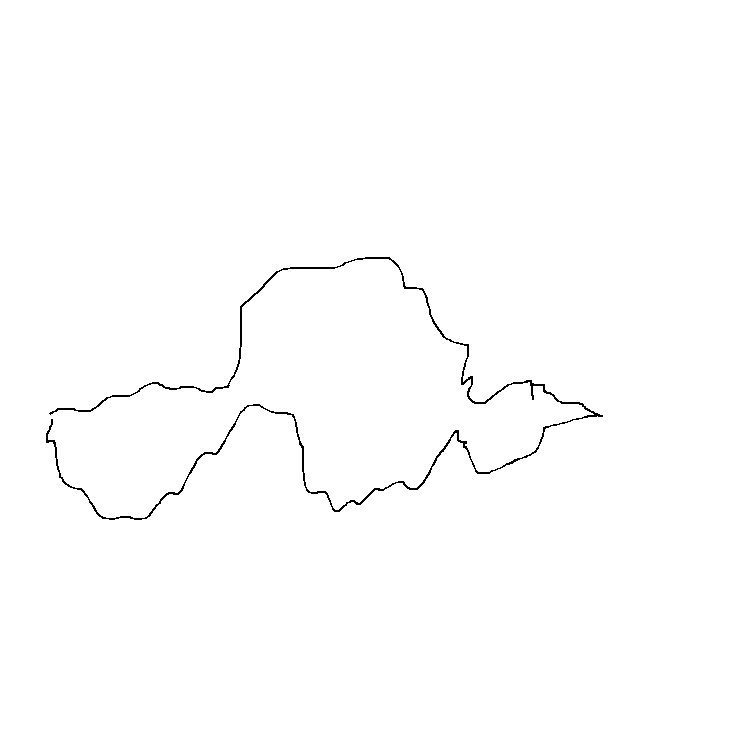} 
    4/9
    & 
    \includegraphics[width=1.05\linewidth]{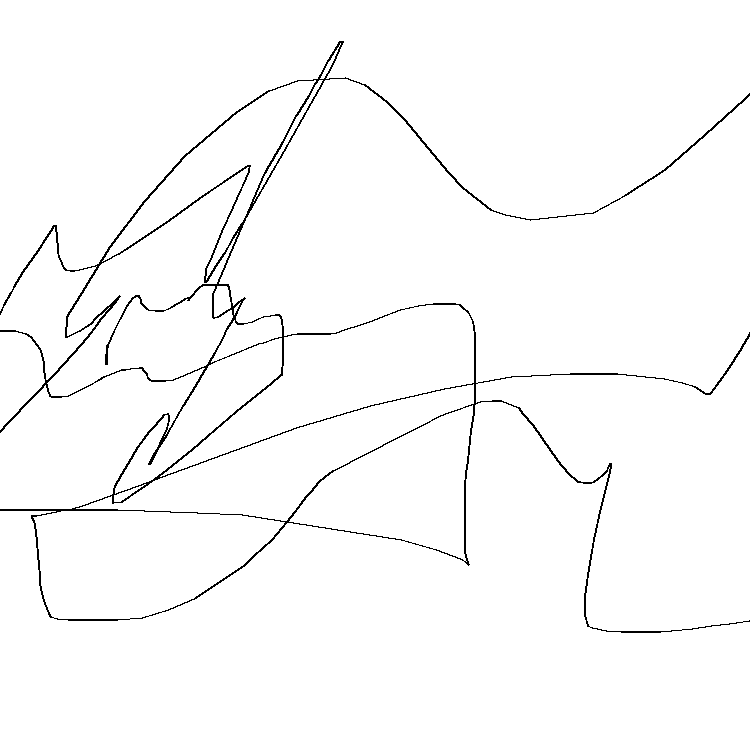}
    10/12
    & 
    \includegraphics[width=1.05\linewidth]{Sketches/pt2snd7B.png}
    8/10
    & 
    \includegraphics[width=1.05\linewidth]{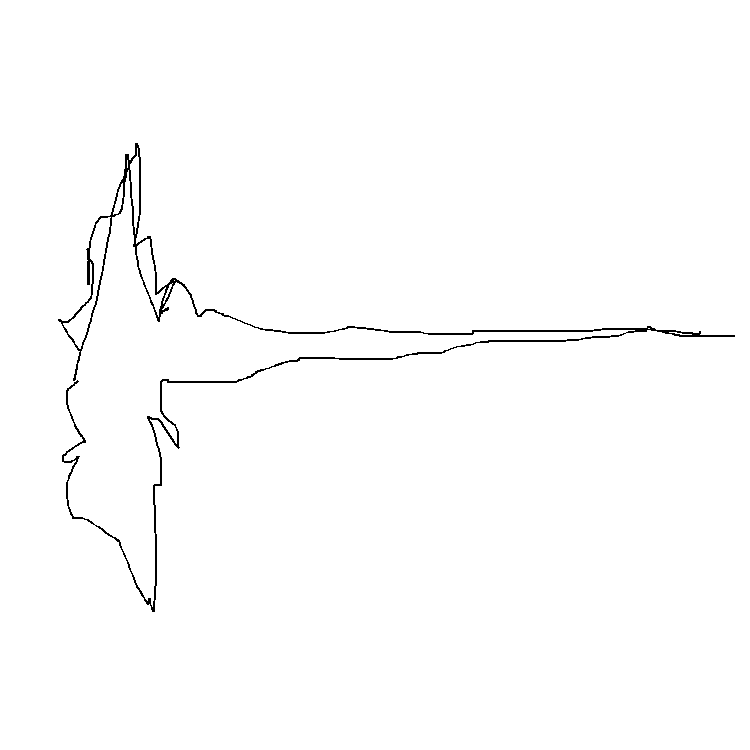} 
    9/12
    &
    \includegraphics[width=1.05\linewidth]{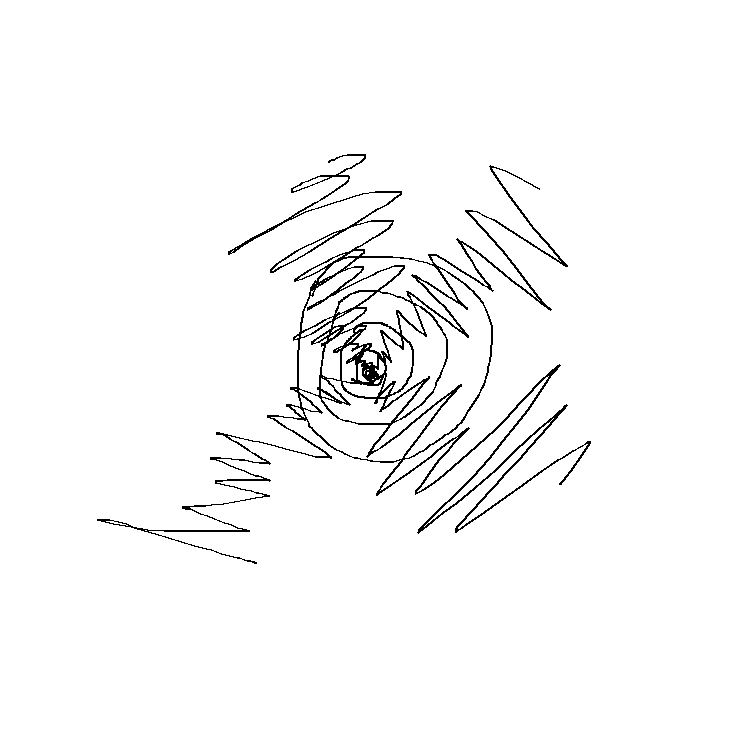}
    6/12
    &
    \includegraphics[width=1.05\linewidth]{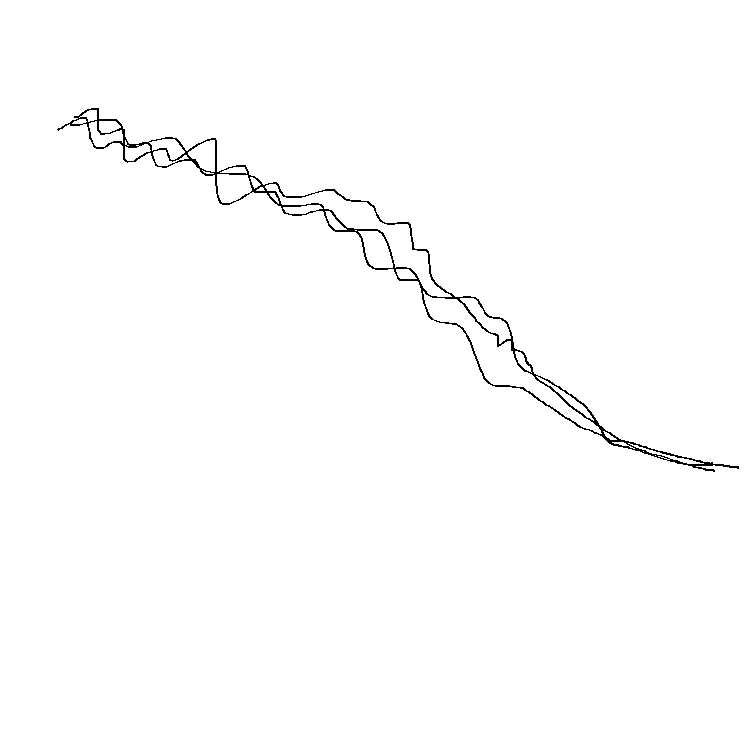}
    \underline{4/9}
  \\
  \hline
  \hline
  Worst
  &
    \includegraphics[width=1.05\linewidth]{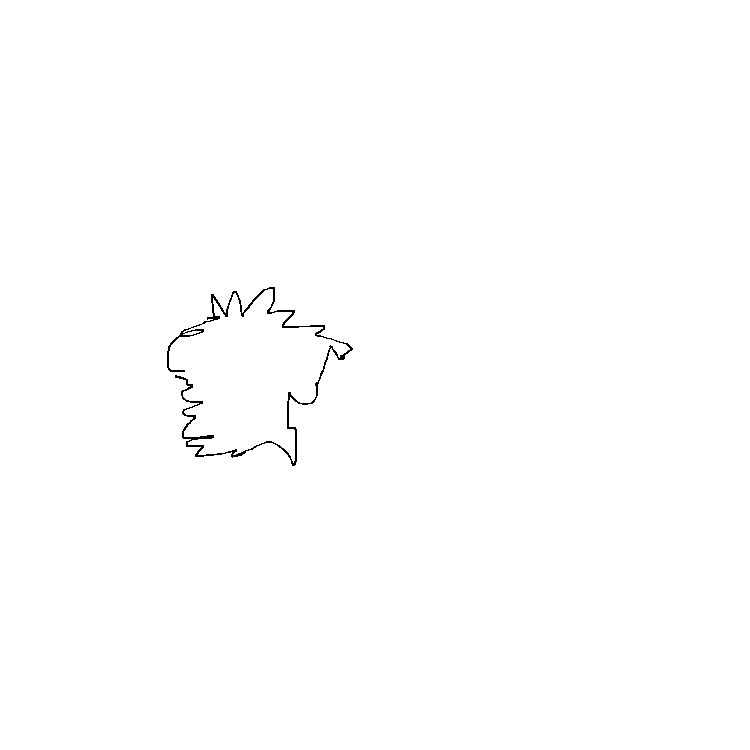} 
    \underline{1/9}
    & 
    \includegraphics[width=1.05\linewidth]{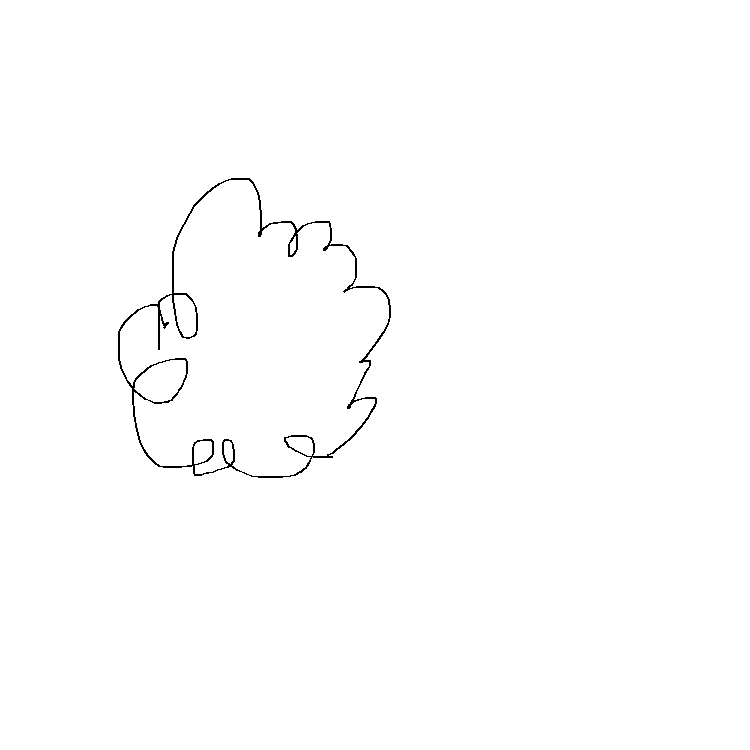} 
    \underline{1/9}
    & 
    \includegraphics[width=1.05\linewidth]{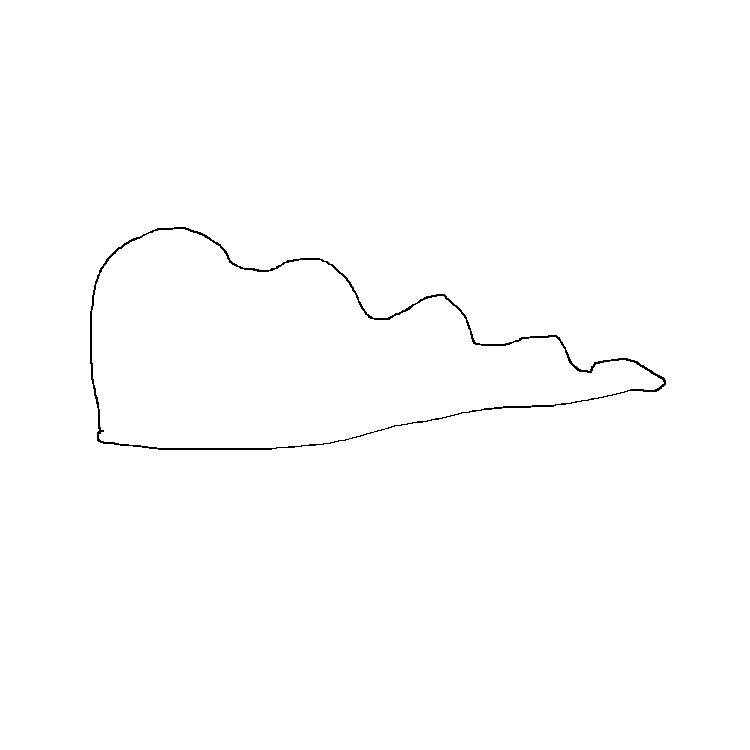} 
    \underline{3/10}
    & 
    \includegraphics[width=1.05\linewidth]{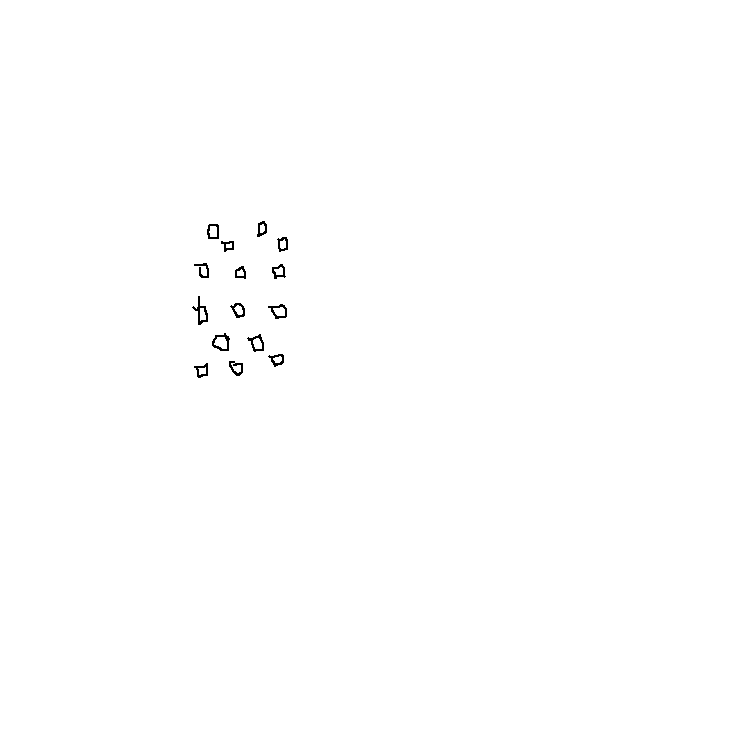}
    0/9
    & 
    \includegraphics[width=1.05\linewidth]{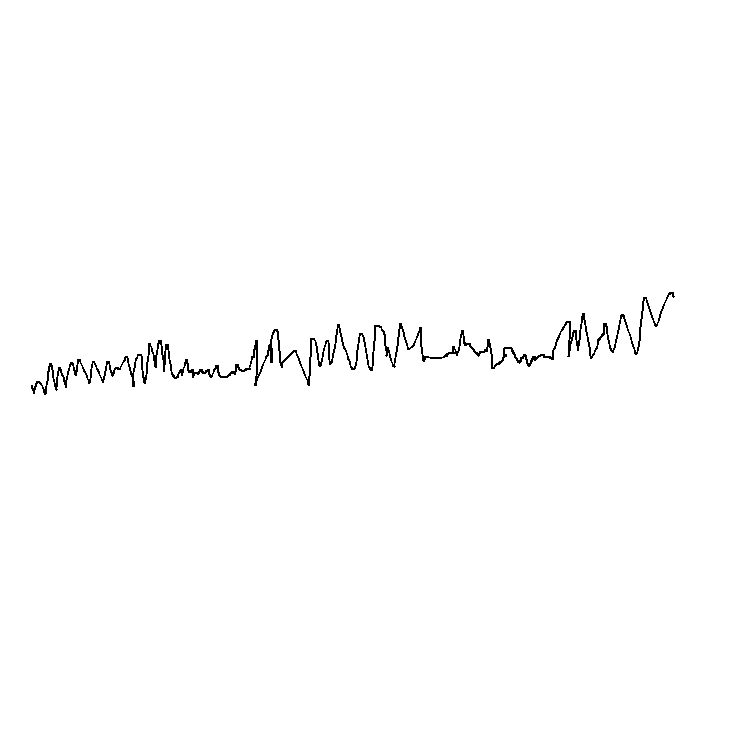} 
    \underline{0/13}
    & 
    \includegraphics[width=1.05\linewidth]{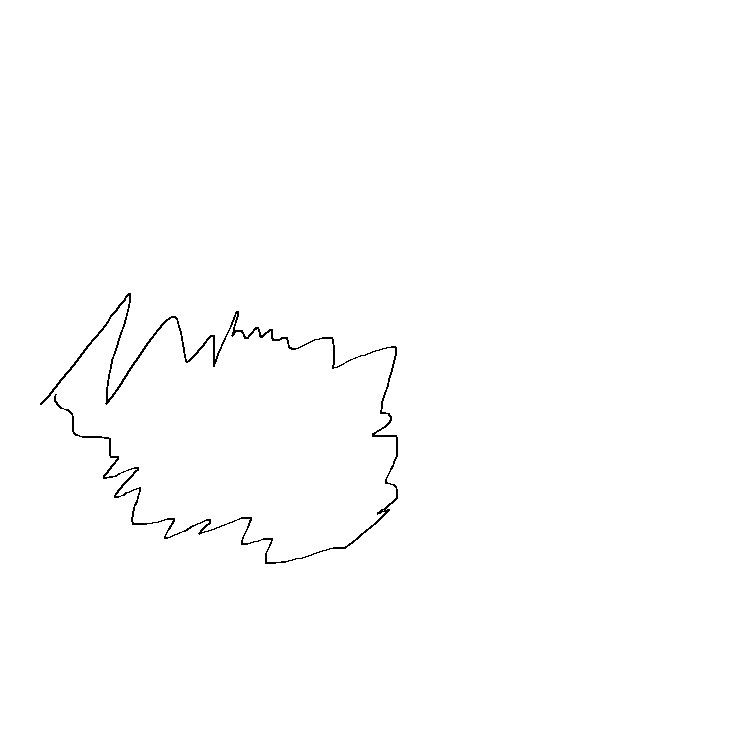} 
    \underline{5/9}
    & 
    \includegraphics[width=1.05\linewidth]{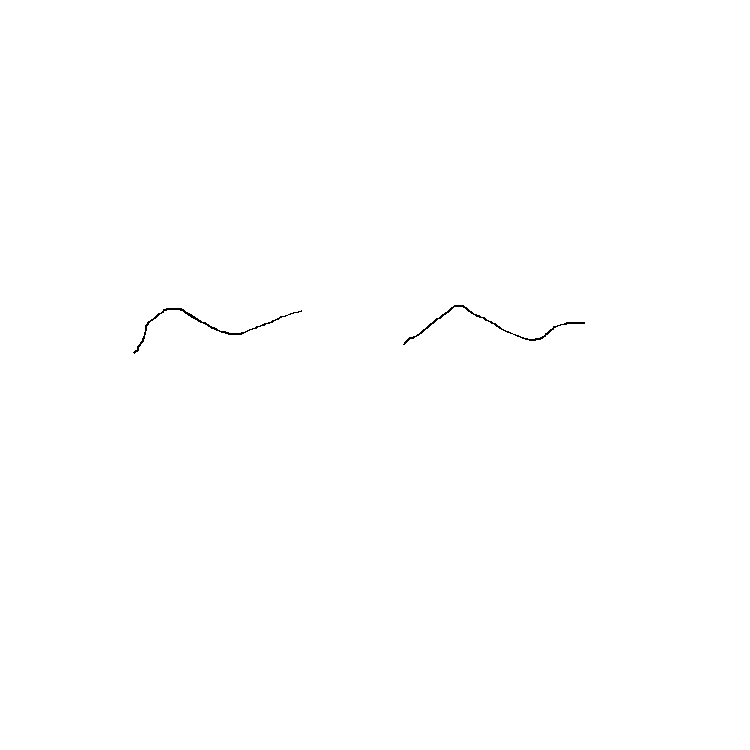} 
    6/10
    & 
    \includegraphics[width=1.05\linewidth]{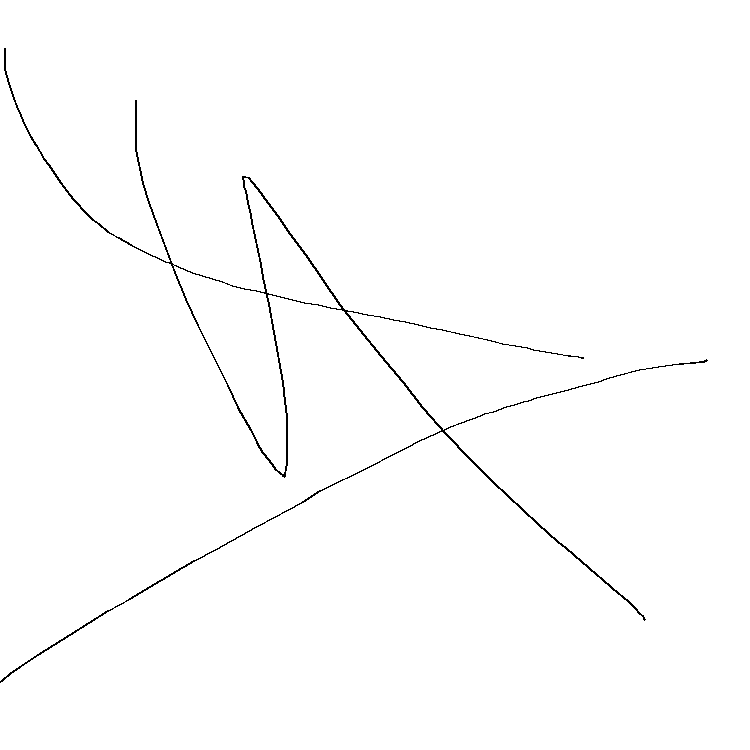}
    2/9
    & 
    \includegraphics[width=1.05\linewidth]{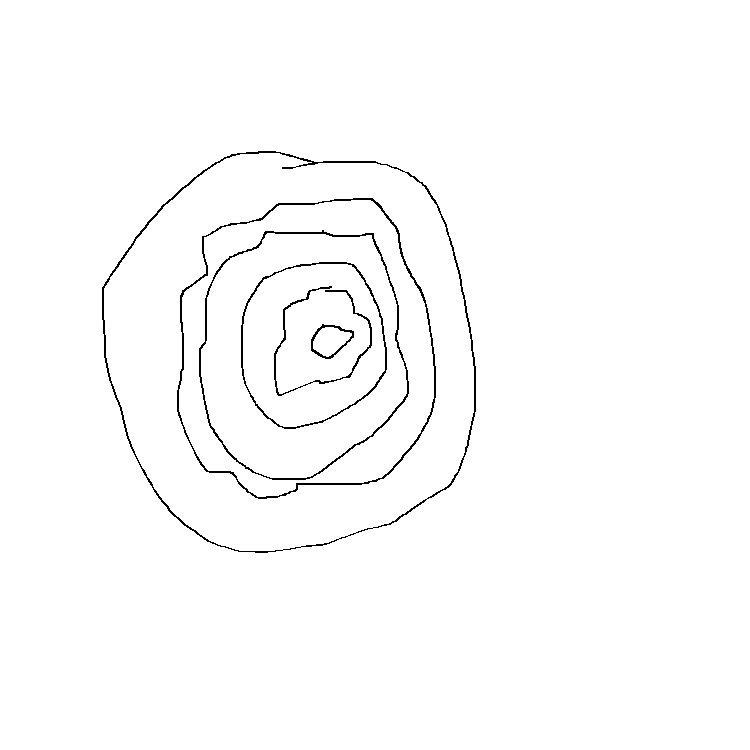}
    \underline{2/10}
    &
    \includegraphics[width=1.05\linewidth]{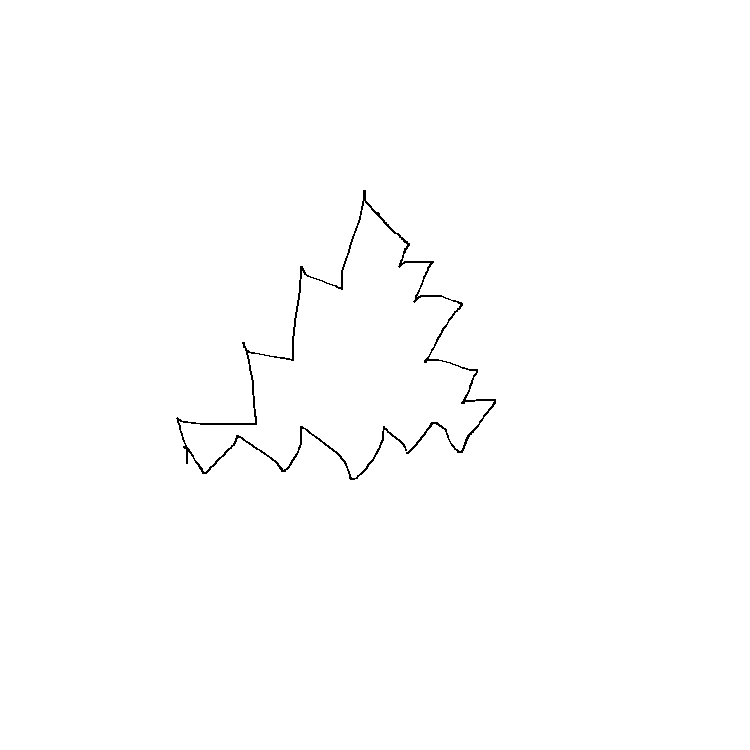}
    \underline{1/10}
    \\
  \hline
 \end{tabular}
\end{center}
\vspace*{-0.05in}
 \caption{Sketches with the best and worst match rates. To reduce the risk of skewing the data through small sample sizes, only sketches were considered that were shown at least 8 times which is the median number of occurrences. Underlined sketches are part of the \textit{Radiating category}.}
 \label{tab:sketches}
\end{table*}

\subsection{Match rate by artist}\label{subsec:hitrate_artist}

The sketching style of an artist appears to have an effect on how well their sketches are matched by participants. Using the sketch categories described in Section \ref{subsec:statistical} a significant correlation could be found between the frequency of sketches in the \textit{Radiating} category and the mean rank for each artist ($\rho$ = -.71, p $<$ .05). This trend can also be observed when comparing categories between the sketches with the highest and lowest match rate for each sound seen in Table \ref{tab:sketches} with \textit{Radiating} occurring 7/10 times for the worst performing sketches, but only 3/10 times for the best performing ones. It has to be noted that from all categories \textit{Radiating} is spread the most evenly across sounds (Cramér's V: .16).

\begin{figure}[h]
\centering
\vspace*{-0.15in}
\includegraphics[width=0.9\columnwidth]{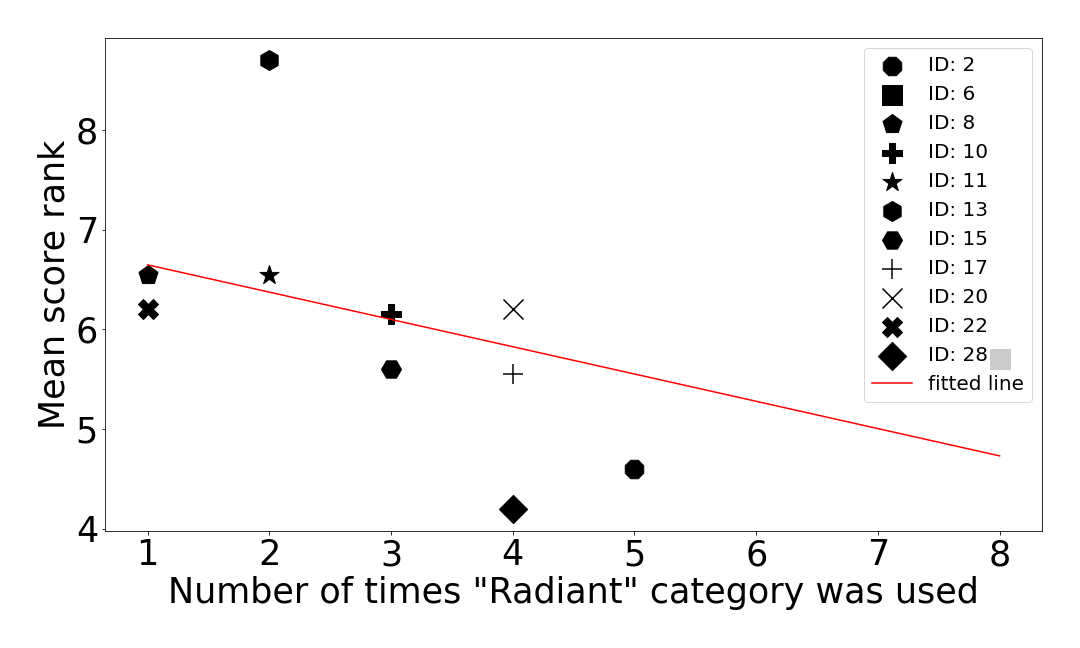}
\vspace*{-0.1in}
\caption{Correlation between match rate rank by artist and prevalence of \textit{Radiating} category in sketches by an artist.}
\label{fig:rank_by_artist}
\end{figure}

\subsection{Study Design Evaluation}\label{subsec:design_evaluation}

\begin{figure}[h]
\centering
\vspace*{-0.15in}
\includegraphics[width=0.9\columnwidth]{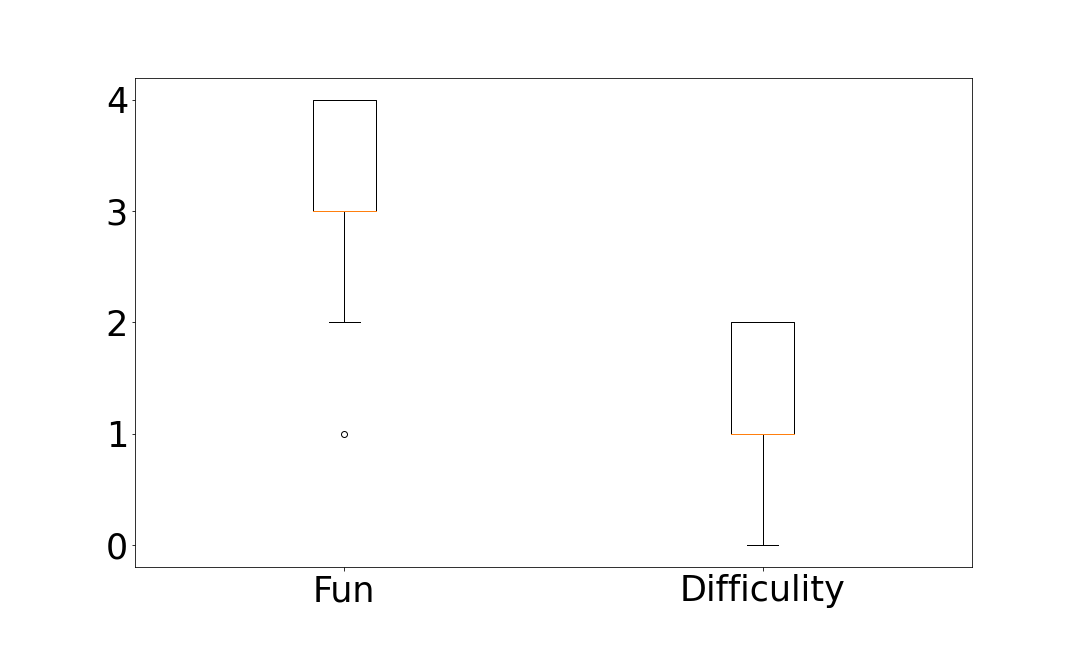}
\vspace*{-0.1in}
\caption{Answers to the questions \textit{I had fun playing the game.} (0-completely disagree to 4-completely agree) and \textit{How difficult was the game?} (0-very difficult to 4-very easy)\label{fig:design_evaluation}}.
\end{figure}

A secondary question of this study was to investigate the effectiveness of the gamified study design. Eighty-two participants were recruited from visitors of the online exhibitions. The age range of participants described in Section~\ref{subsec:participants} indicates that a diverse demographic could be reached. Figure~\ref{fig:design_evaluation} shows that despite finding the task difficult most participants enjoyed completing the study. The responses for \textit{having fun} align with findings by~\cite{harmsGamificationOnlineSurveys2015}. The free-form answers reveal that the score system of the study design was criticised by some participants finding it ``difficult getting a 'score' for a subjective agreement on sound'' (Participant X1) or not being ``sure how you define the 'correct' answer'' (Participants X1). Another point of criticism was the short introduction section with participants saying that ``the instructions went by too quickly'' (Participant X3), the ``interface was tricky to understand'' (Participant X4) and one suggested to ``have like a test sound before the game starts'' (Participant X5).

\section{Discussion}
\label{sec:discussion}
The gamified study design and integration into non-academic exhibitions proved successful at reaching a large and varied participant pool without high recruitment effort. However, spanning over the course of a year, data collection was slow. According to survey responses analysed in Section \ref{subsec:design_evaluation}, participants enjoyed completing the study even though they perceived the task to be difficult. The score system of the design appeared to be confusing to some with survey responses indicating that scores might not be adequate for this task. Some participants seemed to struggle with understanding the interface and the task possibly because the introduction was short and easy to skip. A tutorial or training run could be included to make sure that all participants understand the task. While the integration into an online exhibition might lower initial resistance to participate, it can make it more difficult to motivate participants to complete more tedious tasks like filling in a questionnaire. Figure \ref{fig:scores} shows that a majority of participants did not continue with the post-study survey after submitting their score. Overall, a gamified design presented in a non-academic context might be most suitable for studies that are not time-sensitive and do not require a tightly controlled study environment. 
Results presented in Section \ref{subsec:hitrate} show that participants found the correct sound-sketch matches at significantly higher rate than the random baseline. However, the match rate is notably lower than the results of similar studies that use visual stimuli that were purposefully crafted by a single artist \cite{grillVisualizationPerceptualQualities2012,chen2010thumbnaildj}. This is not unexpected as the sketches used in this study were not produced with the goal to maximise recognition for other people. Instead the results suggest that even sketches that were produced intuitively in a short time can communicate information to identify the sound that they represent. The results further show that match rates differ among different sounds and artists. Sketches from sounds that are more dissimilar to the remaining stimuli are more frequently matched correctly. This suggests that sound-sketches can successfully communicate general sound information but fail to encode more detailed differences in timbre. Against expectation, participants were most likely to select the sketch relating to the most dissimilar sound rather than the next most similar one when not finding the correct match. This could mean that participants adopted a selection process that identifies the ``odd one out'' when three of the four sketches looked similar. Analysis of sketching category frequencies presented in Section \ref{subsec:hitrate_artist} indicate that some sketching styles might be better at conveying information about sound than others, but as noted this could be caused by a more even distribution of some styles across sounds which makes them more likely to be mismatched than a style that is more strongly associated with a specific sound. In addition, outliers to these trends exist for sounds and artists potentially because the quantified measures for sound dissimilarity and sketching style cannot capture all dimensions of human perception. Qualitatively reviewing the results suggests that the perceptual dimension from noisiness/complexity to calmness/simplicity might be encoded especially well in the sketch representations which would explain the high mean match rank for \textit{Noise} and \textit{Piano}.    

\section{Conclusion}
\label{sec:conclusion}
This study shows that graphical sketch representations of sounds can be recognised successfully by human participants if their timbres are sufficiently different from each other, but it appears to be difficult to visually encode timbral differences of similar sounds. In the context of developing a sketch-based sound synthesiser, a deep learning mapping model might be limited to predicting general sound categories even when trained on a larger sound-sketch dataset. Similar studies that reported higher matching rates used purposefully crafted visual stimuli suggesting that introducing guidelines on how to represent different perceptual dimensions could increase a model's accuracy. A different approach could be found in tuning a base model according to an individual user's sketching style.    

\begin{acknowledgments}
EPSRC and AHRC Centre for Doctoral Training in Media and Arts Technology (EP/L01632X/1).
\end{acknowledgments} 


\bibliography{icmc2022template}

\begin{thebibliography}{10}
\providecommand{\url}[1]{#1}
\csname url@samestyle\endcsname
\providecommand{\newblock}{\relax}
\providecommand{\bibinfo}[2]{#2}
\providecommand{\BIBentrySTDinterwordspacing}{\spaceskip=0pt\relax}
\providecommand{\BIBentryALTinterwordstretchfactor}{4}
\providecommand{\BIBentryALTinterwordspacing}{\spaceskip=\fontdimen2\font plus
\BIBentryALTinterwordstretchfactor\fontdimen3\font minus
  \fontdimen4\font\relax}
\providecommand{\BIBforeignlanguage}[2]{{%
\expandafter\ifx\csname l@#1\endcsname\relax
\typeout{** WARNING: IEEEtran.bst: No hyphenation pattern has been}%
\typeout{** loaded for the language `#1'. Using the pattern for}%
\typeout{** the default language instead.}%
\else
\language=\csname l@#1\endcsname
\fi
#2}}
\providecommand{\BIBdecl}{\relax}
\BIBdecl

\bibitem{saitisTimbreSemanticsLens2018}
C.~Saitis, S.~Weinzierl, K.~von Kriegstein, S.~Ystad, and C.~Cuskley, ``Timbre
  semantics through the lens of crossmodal correspondences: A new way of asking
  old questions,'' \emph{Acoustical Science and Technology}, vol.~41, no.~1,
  pp. 365--368, 2020.

\bibitem{siedenburgComparisonApproachesTimbre2016}
K.~Siedenburg, I.~Fujinaga, and S.~McAdams, ``A comparison of approaches to
  timbre descriptors in music information retrieval and music psychology,''
  \emph{Journal of New Music Research}, vol.~45, no.~1, pp. 27--41, 2016.

\bibitem{mcadamsPerceptionMusicalTimbre2016}
S.~McAdams and B.~L. Giordano, ``The perception of musical timbre,'' \emph{The
  Oxford handbook of music psychology}, pp. 72--80, 2009.

\bibitem{martinoSynesthesiaStrongWeak2001}
G.~Martino and L.~E. Marks, ``Synesthesia: Strong and weak,'' \emph{Current
  Directions in Psychological Science}, vol.~10, no.~2, pp. 61--65, 2001.

\bibitem{davisFitnessNamesDrawings1961}
R.~Davis, ``\BIBforeignlanguage{en}{The {{Fitness}} of {{Names}} to
  {{Drawings}}. a {{Cross}}-{{Cultural Study}} in {{Tanganyika}}},''
  \emph{\BIBforeignlanguage{en}{British Journal of Psychology}}, vol.~52,
  no.~3, pp. 259--268, 1961.

\bibitem{taylorPhoneticSymbolismFour1962}
I.~K. Taylor and M.~M. Taylor, ``Phonetic Symbolism in Four Unrelated
  Languages,'' \emph{Canadian Journal of Psychology/Revue canadienne de
  psychologie}, vol.~16, no.~4, pp. 344--356, 1962.

\bibitem{bremnerBoubaKikiNamibia2013}
A.~J. Bremner, S.~Caparos, J.~Davidoff, J.~de~Fockert, K.~J. Linnell, and
  C.~Spence, ``“Bouba” and “Kiki” in Namibia? A remote culture make
  similar shape--sound matches, but different shape--taste matches to
  Westerners,'' \emph{Cognition}, vol. 126, no.~2, pp. 165--172, 2013.

\bibitem{maurerShapeBoubasSound2006}
D.~Maurer, T.~Pathman, and C.~J. Mondloch, ``\BIBforeignlanguage{en}{The Shape
  of Boubas: Sound\textendash Shape Correspondences in Toddlers and Adults},''
  \emph{\BIBforeignlanguage{en}{Developmental Science}}, vol.~9, no.~3, pp.
  316--322, 2006.

\bibitem{adeliAudiovisualCorrespondenceMusical2014}
M.~Adeli, J.~Rouat, and S.~Molotchnikoff, ``Audiovisual correspondence between
  musical timbre and visual shapes,'' \emph{Frontiers in human neuroscience},
  vol.~8, p. 352, 2014.

\bibitem{lobbers2021sketching}
S.~L{\"o}bbers, M.~Barthet, and G.~Fazekas, ``Sketching sounds: an exploratory
  study on sound-shape associations,'' in \emph{International Computer Music
  Conference (ICMC)}, 2021, p.~6.

\bibitem{kolhoffMusicIconsProcedural2006}
P.~Kolhoff, J.~Preub, and J.~Loviscach, ``Music {{Icons}}: {{Procedural
  Glyphs}} for {{Audio Files}},'' in \emph{2006 19th {{Brazilian Symposium}} on
  {{Computer Graphics}} and {{Image Processing}}}, Oct. 2006, pp. 289--296.

\bibitem{chen2010thumbnaildj}
Y.-X. Chen and R.~Kl{\"u}ber, ``ThumbnailDJ: Visual Thumbnails of Music
  Content.'' in \emph{11th International Society for Music Information
  Retrieval Conference (ISMIR)}, 2010, pp. 565--570.

\bibitem{wan2019towards}
C.-H. Wan, S.-P. Chuang, and H.-Y. Lee, ``Towards audio to scene image
  synthesis using generative adversarial network,'' in \emph{ICASSP 2019-2019
  IEEE International Conference on Acoustics, Speech and Signal Processing
  (ICASSP)}.\hskip 1em plus 0.5em minus 0.4em\relax IEEE, 2019, pp. 496--500.

\bibitem{ishibashi2020investigating}
T.~Ishibashi, Y.~Nakao, and Y.~Sugano, ``Investigating audio data visualization
  for interactive sound recognition,'' in \emph{Proceedings of the 25th
  International Conference on Intelligent User Interfaces}, 2020, pp. 67--77.

\bibitem{grillVisualizationPerceptualQualities2012}
T.~Grill and A.~Flexer, ``\BIBforeignlanguage{en}{Visualization of Perceptual
  Qualities in Textural Sounds},'' in
  \emph{\BIBforeignlanguage{en}{International Computer Music Conference
  (ICMC)}}, 2012, p.~8.

\bibitem{kneesSearchingAudioSketching2016}
P.~Knees and K.~Andersen, ``Searching for audio by sketching mental images of
  sound: A brave new idea for audio retrieval in creative music production,''
  in \emph{Proceedings of the 2016 ACM on International Conference on
  Multimedia Retrieval}, 2016, pp. 95--102.

\bibitem{engeln2021similarity}
L.~Engeln, N.~L. Le, M.~McGinity, and R.~Groh, ``Similarity Analysis of Visual
  Sketch-based Search for Sounds,'' in \emph{Audio Mostly 2021}, 2021, pp.
  101--108.

\bibitem{haNeuralRepresentationSketch2017}
D.~Ha and D.~Eck, ``\BIBforeignlanguage{en}{A {{Neural Representation}} of
  {{Sketch Drawings}}},'' \emph{\BIBforeignlanguage{en}{arXiv:1704.03477 [cs,
  stat]}}, May 2017.

\bibitem{hamariDoesGamificationWork2014}
J.~Hamari, J.~Koivisto, and H.~Sarsa, ``Does {{Gamification Work}}?
  \textendash{} {{A Literature Review}} of {{Empirical Studies}} on
  {{Gamification}},'' in \emph{2014 47th {{Hawaii International Conference}} on
  {{System Sciences}}}, Jan. 2014, pp. 3025--3034.

\bibitem{rappStrengtheningGamificationStudies2019}
A.~Rapp, F.~Hopfgartner, J.~Hamari, C.~Linehan, and F.~Cena,
  ``\BIBforeignlanguage{en}{Strengthening Gamification Studies: {{Current}}
  Trends and Future Opportunities of Gamification Research},''
  \emph{\BIBforeignlanguage{en}{International Journal of Human-Computer
  Studies}}, vol. 127, pp. 1--6, Jul. 2019.

\bibitem{harmsGamificationOnlineSurveys2015}
J.~Harms, S.~Biegler, C.~Wimmer, K.~Kappel, and T.~Grechenig, ``Gamification of
  online surveys: Design process, case study, and evaluation,'' in \emph{IFIP
  Conference on Human-Computer Interaction}.\hskip 1em plus 0.5em minus
  0.4em\relax Springer, 2015, pp. 219--236.

\bibitem{duffy2018makes}
S.~Duffy and M.~Pearce, ``What makes rhythms hard to perform? An investigation
  using Steve Reich’s Clapping Music,'' \emph{Plos one}, vol.~13, no.~10, p.
  e0205847, 2018.

\bibitem{martin2020data}
C.~P. Martin and J.~Torresen, ``Data-Driven Analysis of Tiny Touchscreen
  Performance with MicroJam,'' \emph{Computer Music Journal}, vol.~43, no.~4,
  pp. 41--57, 2020.

\bibitem{morschheuserHowGamifyMethod2017}
B.~Morschheuser, J.~Hamari, K.~Werder, and J.~Abe,
  ``\BIBforeignlanguage{en}{How to {{Gamify}}? {{A Method For Designing
  Gamification}}},'' in \emph{\BIBforeignlanguage{en}{Hawaii {{International
  Conference}} on {{System Sciences}}}}, 2017.

\bibitem{mullensiefenGoldsmithsMusicalSophistication}
D.~M{\"u}llensiefen, B.~Gingras, J.~Musil, and L.~Stewart, ``The musicality of
  non-musicians: an index for assessing musical sophistication in the general
  population,'' \emph{PloS one}, vol.~9, no.~2, p. e89642, 2014.

\bibitem{pearceModellingTimbralHardness2019}
A.~Pearce, T.~Brookes, and R.~Mason, ``\BIBforeignlanguage{en}{Modelling
  {{Timbral Hardness}}},'' \emph{\BIBforeignlanguage{en}{Applied Sciences}},
  vol.~9, no.~3, p. 466, 2019.

\bibitem{braunUsingThematicAnalysis2006}
V.~Braun and V.~Clarke, ``\BIBforeignlanguage{en}{Using Thematic Analysis in
  Psychology},'' \emph{\BIBforeignlanguage{en}{Qualitative Research in
  Psychology}}, vol.~3, no.~2, pp. 77--101, 2006.

\end{thebibliography}

\end{document}